\documentclass[aps,prl,preprint,nopacs,superscriptaddress]{revtex4}

\usepackage{graphicx}
\usepackage{verbatim}
\usepackage{mathrsfs}
\usepackage{gensymb}
\usepackage{color}
\usepackage{ulem}
\usepackage{pdfpages}

\usepackage{amsmath,amsfonts,amssymb}
\usepackage{graphicx}
\usepackage{amsmath,cleveref,amssymb,amsfonts,graphicx,multirow,color,bm,textcomp,mathtools}
\usepackage{paralist}
\usepackage{srcltx}
\usepackage[all,cmtip]{xy}
\usepackage{mathrsfs}
\usepackage{srcltx,soul,subfigure}
\usepackage{threeparttable,dsfont,bbm}

\usepackage{hyperref}
\usepackage{hyperref}
\usepackage{color}

\crefname{equation}{Eq.}{Eqs.}
\crefname{figure}{Fig.}{Figs.}

\newcommand{\ket}[1]{| #1 \rangle}

\newcommand{\bee}{\begin{eqnarray}}
\newcommand{\ee}{\end{eqnarray}}
\newcommand{\bma}{\begin{pmatrix}}
\newcommand{\ema}{\end{pmatrix}}
\newcommand{\balig}{\begin{align}}
\newcommand{\ealig}{\end{align}}

\newcommand{\ba}{\begin{align}}
\newcommand{\ea}{\end{align}}

\newcommand{\ignore}[1]{}

\def\3{2.8in}    
\def\2{2.5in}
\def\4{3.0in}\def \beq {\begin{equation}}
\def \eeq {\end{equation}}
\begin{document}

\title{Realisation of Symmetry Enforced Two-Dimensional Dirac Fermions in Nonsymmorphic $\alpha$-Bismuthene}

\author{Pawel~J. Kowalczyk}
\affiliation {Department of Solid State Physics, Faculty of Physics and Applied Informatics, University of Lodz,
90-236 Lodz, Pomorska 149/153, Poland}

\author{Simon~A. Brown}
\affiliation {The MacDiarmid Institute for Advanced Materials and Nanotechnology, School of Physical and Chemical Sciences,
University of Canterbury, Private Bag 4800, Christchurch 8140, New Zealand}

\author{Tobias~Maerkl}
\affiliation {The MacDiarmid Institute for Advanced Materials and Nanotechnology, School of Physical and Chemical Sciences,
University of Canterbury, Private Bag 4800, Christchurch 8140, New Zealand}

\author{Qiangsheng Lu}
\affiliation {Department of Physics and Astronomy, University of Missouri, Columbia, Missouri 65211, USA}

\author{Ching-Kai Chiu}
\affiliation {Kavli Institute for Theoretical Sciences, University of Chinese Academy of Sciences, Beijing 100190, China}

\author{Ying Liu}
\affiliation {Research Laboratory for Quantum Materials, Singapore University of Technology and Design, Singapore 487372, Singapore}

\author{Shengyuan A. Yang}
\affiliation {Research Laboratory for Quantum Materials, Singapore University of Technology and Design, Singapore 487372, Singapore}

\author{Xiaoxiong~Wang}
\affiliation {College of Science, Nanjing University of Science and Technology, Nanjing 210094, China}

\author{Ilona~Zasada}
\affiliation {Department of Solid State Physics, Faculty of Physics and Applied Informatics, University of Lodz,
90-236 Lodz, Pomorska 149/153, Poland}

\author{Francesca~Genuzio}
\affiliation {Elettra - Sincrotrone Trieste S.C.p.A., Basovizza, I-34149 Trieste, Italy}

\author{T.~Onur~Mente\c{s}}
\affiliation {Elettra - Sincrotrone Trieste S.C.p.A., Basovizza, I-34149 Trieste, Italy}

\author{Andrea~Locatelli}
\affiliation {Elettra - Sincrotrone Trieste S.C.p.A., Basovizza, I-34149 Trieste, Italy}

\author{T.-C.~Chiang}
\affiliation {Department of Physics, University of Illinois at Urbana-Champaign, 1110 West Green Street, Urbana, Illinois 61801-3080, USA}
\affiliation {Frederick Seitz Materials Research Laboratory, University of Illinois at Urbana-Champaign, 104 South Goodwin Avenue, Urbana, Illinois 61801-2902, USA}

\author{Guang~Bian}
\affiliation {Department of Physics and Astronomy, University of Missouri, Columbia, Missouri 65211, USA}

\pacs{}

\begin{abstract}

Two-dimensional (2D) Dirac-like electron gases have attracted tremendous research interest ever since the discovery of free-standing graphene \cite{Novoselov2005, Kim2005, Kane2010}. The linear energy dispersion and non-trivial Berry phase play the pivotal role in the remarkable electronic, optical, mechanical and chemical properties of 2D Dirac materials \cite{Castro2009}. The known 2D Dirac materials are gapless only within certain approximations, for example, in the absence of SOC. Here we report a route to establishing robust Dirac cones in 2D materials with nonsymmorphic crystal lattice. The nonsymmorphic symmetry enforces Dirac-like band dispersions around certain high-symmetry momenta in the presence of SOC \cite{Kane2016, Po2017}. Through $\mu$-ARPES measurements we observe Dirac-like band dispersions in $\alpha$-bismuthene. The nonsymmorphic lattice symmetry is confirmed by $\mu$-LEED and STM. Our first-principles simulations and theoretical topological analysis demonstrate the correspondence between nonsymmorphic symmetry and Dirac states. This mechanism can be straightforwardly generalized to other nonsymmorphic materials. The results open the door for the search of symmetry enforced Dirac fermions in the vast uncharted world of nonsymmorphic 2D materials. 

\end{abstract}

\maketitle

The discovery of graphene and topological insulators has stimulated enormous interest in two dimensional electron gas with linear band dispersion\cite{Novoselov2005, Kim2005, Kane2010}. The vanishing effective mass and non-zero Berry phase of Dirac fermion-like states give rise to many remarkable physical properties such as extremely high mobilities and zero-energy Landau levels\cite{Castro2009}. The two dimensional Dirac cones are generally fragile against perturbations and so various gapped electronic phases are observed\cite{Zhou2008}. For example, the weak spin-orbit coupling (SOC) can make graphene a quantum spin Hall insulator.  Gapless 2D Dirac fermions can be found on the surfaces of 3D topological insulators~\cite{Kane2010}. However, when the dimension of the 3D systems is reduced, a tunneling energy gap opens at the Dirac point due to the hybridization of the surface states on the opposite surfaces. Therefore, it is a highly challenging task to find gapless Dirac fermions in intrinsic 2D materials. So far, only a few 2D materials have been theoretically predicted to possess Dirac states, including graphene\cite{Novoselov2005, Kim2005}, silicene\cite{silicene}, germanene\cite{germanene}, and some artificial structures\cite{graphynes2012, Oganov2014, Ko2012, Picket2009}. Among them, only graphene has been experimentally proven to possess Dirac cones since the spin-orbit gap is negligibly small.

The reason for the rarity of 2D Dirac materials can be seen from a simple co-dimension analysis. The general Hamiltonian of 2D Dirac states is described by $\alpha\sigma_{x}k_{x}+\beta\sigma_{y}k_{y}$, where $\sigma_{i}$ and $k_{i}$ are Pauli matrices and momentum components, and $\alpha$ and $\beta$ are constants. The Dirac cone can be easily gapped by any perturbations in the form of $m\sigma_z$ and the resulted gapped system is energetically favored given that the Fermi level is in the gap. The perturbation can be from the intrinsic SOC of the system as in graphene, or from lattice relaxations which reduce the surface energy of the 2D material \cite{Bian-PRB90-195409-14}. In order to achieve truly gapless Dirac dispersion, geometrical or topological constraints are required to eliminate the gap term. Previous theoretical studies of the relation between lattice symmetry and Dirac states, suggested that nonsymmorphic symmetries can enforce Dirac-like band structures \cite{Young2015, Guan2017, Kane2016, Wieder2018, Po2017}. The key idea is that the operator algebra involving nonsymmorphic symmetries allows only nontrivial irreducible representations at certain high-symmetry points of the Brillouin zone \cite{Kane2016, Po2017}. However, to date, a material realization of symmetry enforced Dirac fermions in 2D nonsymmorphic materials is still elusive. 

In this work, we report the existence of 2D Dirac fermions in $\alpha$-bismuthene (``$\alpha$-Bi" for short). The Dirac band structure is observed by our micro-angle-resolved photoemission ($\mu$-ARPES) experiment.  The Dirac cone is protected by the glided mirror of the nonsymmorphic $\alpha$-bismuthene lattice and show that this concept can be generalized to other 2D materials with nonsymmorphic layer groups. The Dirac points are located at certain high-symmetry momentum points which are entirely determined by the symmetry of the lattice. This correspondence of Dirac states to the lattice symmetry opens the way to discovery of a range of new 2D Dirac materials. In this paper, we first report our experimental results on  $\alpha$-bismuthene. Then we will present a theoretical analysis of the 2D Dirac cone in this material as well as the first-principles band structure of bismuthene films. Finally, we discuss a guiding rule for the search of 2D Dirac materials.



\begin{figure}
\includegraphics[width=1\linewidth]{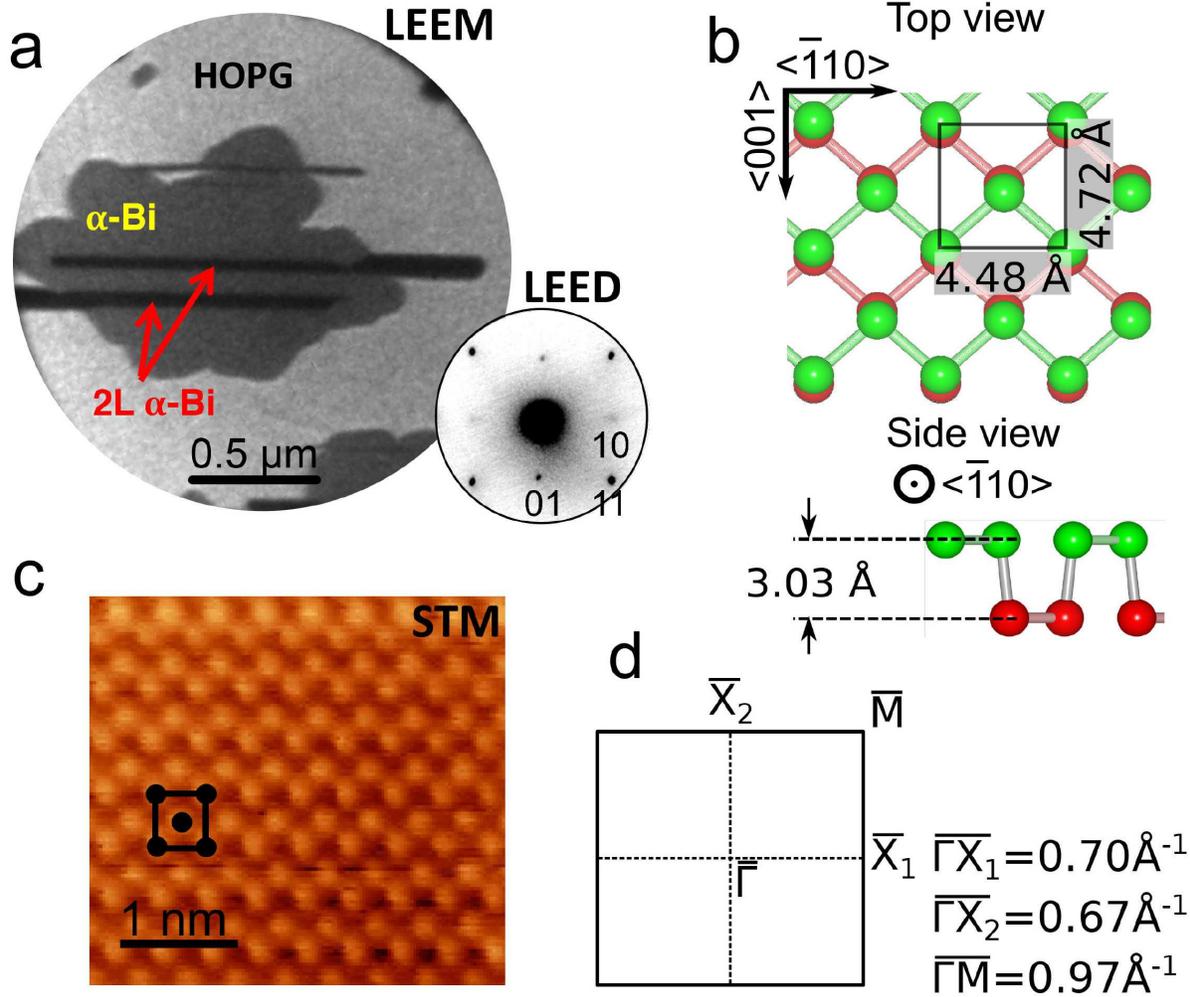}[H]
\caption{
\label{fig:model}
\textbf{Lattice structure of $\alpha$-bismuthene.} \textbf{a,} LEEM image of the $\alpha$-Bi island. $\mu$-LEED pattern recorded on $\alpha$-Bi island is shown in the inset. \textbf{b,} Top and side views of $\alpha$-Bi lattice structure. The structure belongs to the $\#$42 layer group $pman$. Dimension of the unit cell after full DFT optimization is shown \textbf{b}. \textbf{c,}, STM image of $\alpha$-Bi with atomic resolution. \textbf{d,} The Brillouin zone of $\alpha$-Bi.
}%
\end{figure}

 In our experiment, $\alpha$-Bi was grown on HOPG substrates under an ultrahigh vacuum environment. In Fig.~\ref{fig:model}\textbf{a}, a typical LEEM image recorded on $\alpha$-bismuthene ($\alpha$-Bi) is shown. $\alpha$-Bi is relatively well characterized and consists of 2-monolayer thick butterfly-like bases with black phosphorus-like crystallographic structure and the (110) plane (rhombohedral indices) parallel to the substrate\cite{Kowalczyk-NL13-43-13, PhysRevB.73.205424} (see Fig.~\ref{fig:model}\textbf{b}).  On top of $\alpha$-Bi, additional stripes of $\alpha$-Bi are usually observed. These can be seen in LEEM due to quantum oscillations in electron reflectivity for electron energies below 10~eV (as a result of the long inelastic mean free path). In our experiments, we found that highest height contrast for Bi on HOPG is obtained for electron energy equal to 8~eV (see Fig.~\ref{fig:model}\textbf{a}). 

By using $\mu$-LEED (aperture size 500~nm),
we confirm the crystallographic structure of these islands is of black-phosphorus type rather than, for example, possible hexagonal structures\cite{Koroteev2008,Hofmann-PiSS81-191-06} (see top inset in Fig.~\ref{fig:model}\textbf{a} recorded on island shown in Fig.~\ref{fig:model}\textbf{a}). Close inspection of the $\mu$-LEED pattern reveals the very weak intensity of the (10) spots, which is the result of glide-line symmetry along the $\langle \bar{1} 1 0 \rangle$ direction\cite{Lizzit-SS603-3222-09} (rhombohedral indices, see Fig.~\ref{fig:model}\textbf{b}). Detailed analysis on the LEED intensity pattern indicates that the two atoms in each layer in the unit cell are at nearly the same height; the surface buckling is small, with a deformation $\Delta z$ less than 0.04 \AA, please see the supplementary information for a detailed discussion. The geometry of the unit cell from the LEED fitting is shown in Fig.~\ref{fig:model}\textbf{b}. This structure belongs to the $\#$42 layer group ($pman$). The dimensions of the unit cell from STM measurements are $(4.5 \pm 0.2)\times(4.8 \pm 0.2)$~\AA$^2$; see the inset in Fig.~1{\bf a}. The two-atom surface unit cell from STM is consistent with the LEED result. This is further supported by our fully relaxed first-principles calculations (in the frame work of density functional theory (DFT)) performed for a freestanding film yielding a unit cell of $4.48 \times 4.72$~\AA$^2$ (see Fig.~\ref{fig:model}\textbf{b}). Note that in this structure there is no buckling within the uncertainties of the calculations, that is, the atoms in each layer are exactly parallel to the surface plane; please see a detailed discussion in the supplementary information. The Brillouin zone is plotted in Fig.~1\textbf{c}, in which $\bar{\Gamma}$-$\bar{\mathrm{X}}_1$ is along the $\langle\bar{1}10\rangle$ direction and $\bar{\Gamma}$-$\bar{\mathrm{X}}_2$ along the $\langle001\rangle$ direction.

\begin{figure}
\includegraphics[width=1\linewidth]{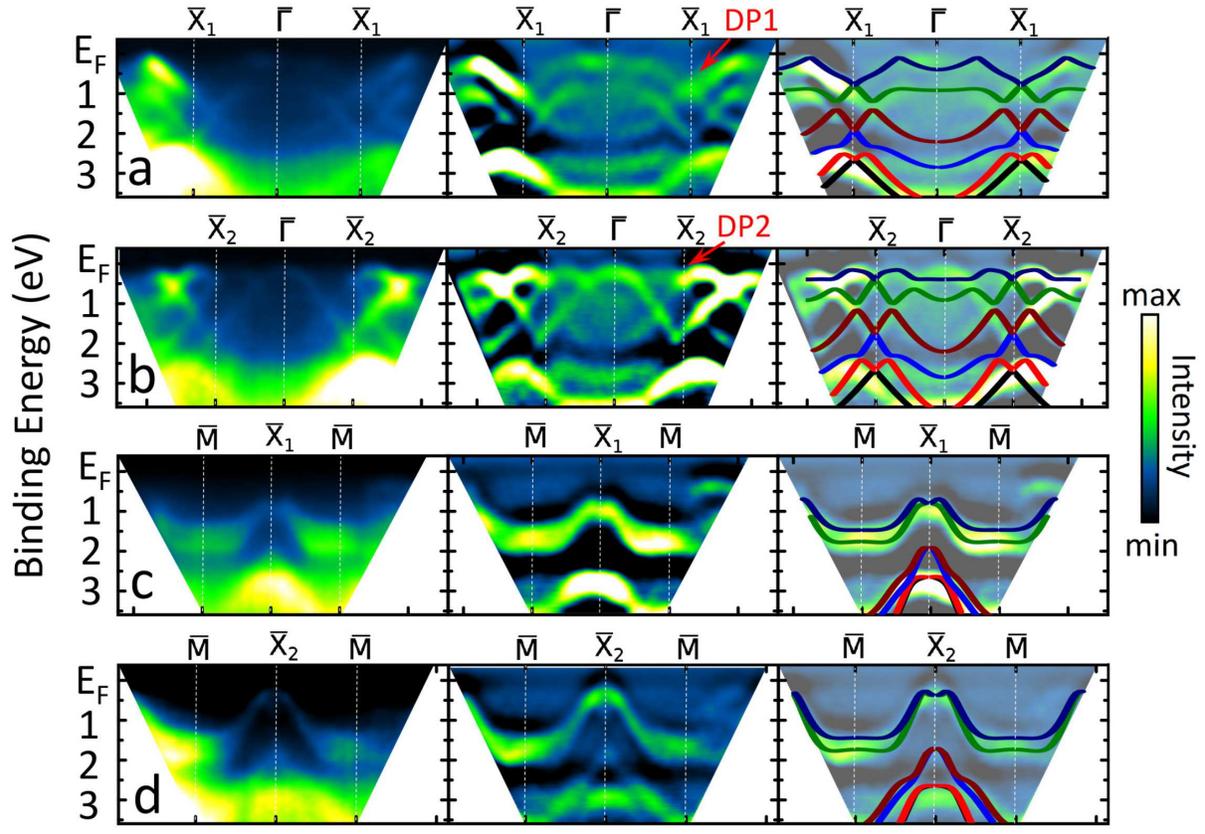}[H]
\caption{
\label{fig:arpes}
\textbf{$\mu$-ARPES band structure taken from a single island of $\mathbf{\alpha}$-Bi.} \textbf{a}-\textbf{d,}  The ARPES band mapping taken at photon energy 27.9 eV along different high-symmetry directions as indicated in each subfigure. The raw data is shown in the left column, second-derivative enhanced data in the middle, and ARPES result overlaid with first-principles bands in the right column.}%
\end{figure}

In order to understand the electronic properties of $\alpha$-Bi, we performed $\mu$-ARPES measurements using a spectroscopic photoemission and low-energy electron microscope (SPELEEM)\cite{Mentes-BJN5-1873-14} on the island shown in Fig.~\ref{fig:model}\textbf{a}. When operated in diffraction imaging mode, the SPELEEM microscope can record the ARPES pattern up to $k_\parallel$ such that the first and a large portion of second Brillouin zones of $\alpha$-Bi are imaged. Results of $\mu$-ARPES experiments are compared with DFT calculations in Fig.~\ref{fig:arpes} along high-symmetry directions of the surface Brillouin zone. Second derivative procedures are used to enhance the visibility of the band features (smoothing and moving average used before 2\textsuperscript{nd} derivative calculation), as in the central column of Fig.~\ref{fig:arpes}. $\alpha$-Bi is a semiconductor with Fermi level barely touching the top of valence band as shown in the ARPES spectra. The most prominent feature of valence band is that there exist band crossings at $\bar{\mathrm{X}}_1$ and $\bar{\mathrm{X}}_2$. A band degeneracy occurs for every band crossing at these two high-symmetry points. Here we focus on the band crossings at 0.7 eV and 0.4 eV (denoted by ``DP1" and ``DP2", respectively, in Fig.~2) as examples and present a detailed analysis of the band dispersion. The conclusions from this analysis apply to every pair of bands that cross at $\bar{\mathrm{X}}_1$ and $\bar{\mathrm{X}}_2$. Overall, the agreement between theory and experiment is remarkably good considering the ARPES spectrum is taken from a single $\alpha$-Bi island with size about 1~$\mu$m$\times$1~$\mu$m. We note that there are no spectral features ascribed to the HOPG substrate, since in the vicinity of Fermi level, HOPG bands are located farther way from the center of the Brillouin zone.

The agreement between experiment and theory is seen very clearly in the plots of iso-energy contours, see Fig.~3\textbf{a}. The calculated contours are plotted on top of experimental images for comparison.
In the first Brillouin zone, there are four small hole pockets at 0.2 eV, two of which are in between $\bar{\Gamma}$ and $\bar{\mathrm{X}}_1$ and the other two close to $\bar{\mathrm{X}}_2$. In the vicinity of $\bar{\mathrm{X}}_2$, the two hole pockets grow into a star-shaped contour as the binding energy goes from 0.2 eV to 0.4 eV. The center of the star corresponds to the Dirac point DP2 at $\bar{\mathrm{X}}_2$. The two pockets in between $\bar{\Gamma}$ and $\bar{\mathrm{X}}_1$ also grow larger as the binding energy increases, and at 0.7 eV, they touch the pockets in the second Brillouin zone at $\bar{\mathrm{X}}_1$, forming the Dirac point DP1.  These pockets eventually merge together into squarish contours for higher binding energy values. This evolution of band contours is consistent with the DFT simulations. The DFT band structure in Fig.~3\textbf{b} demonstrates the band crossings DP1 and DP2. Figs.~3\textbf{c} and 3\textbf{d} show 3D representations of the bands obtained from ARPES and DFT calculation, respectively. To compare with the experimental result, the DFT bands are smeared out by 0.5 eV. All main features in the ARPES data are well reproduced by DFT calculations, strongly suggesting the existence of nonsymmorphic Dirac states. In particular, the band crossing features marked by arrows are nearly identical, which indicates that a Dirac state indeed exists at $\bar{\mathrm{X}}_1$. We note that the Dirac bands at DP1 and DP2 are anisotropic, especially along $\bar{\mathrm{X}}_2$-$\bar{\mathrm{M}}$ direction. The linear dispersion can been seen only in the close vicinity of $\bar{\mathrm{X}}_1$ and $\bar{\mathrm{X}}_2$, as shown in Figs.~4\textbf{a} and 4\textbf{b}.

According to the ARPES and first-principles results, there exist Dirac cones at $\bar{\mathrm{X}}_1=(\pi,0)$ and $\bar{\mathrm{X}}_2=(0,\pi)$ of the Brillouin zone (for simplicity, the lengths are measured in units of the lattice constants $a_x$ and $a_y$ along $\langle\bar{1}10\rangle$ and $\langle001\rangle$ directions, respectively.).  We now show that these band degeneracies are protected by the nonsymmorphic lattice symmetry. $\alpha$-Bi is nonmagnetic and centrosymmetric, so the time reversal $(T)$ and inversion $(P)$ symmetries are preserved. (Note again the lack of buckling.) The space-time inversion symmetry $PT$ leads to the two-fold Kramers degeneracy of each band in the Brillouin zone in the presence of SOC.  Therefore, the band degeneracy is 4 for the band crossing points at $\bar{\mathrm{X}}_1$ and $\bar{\mathrm{X}}_2$.  The lattice of $\alpha$-bismuthene belongs to the \#~42 layer group  ($pman$) which is described by the three generators:
\begin{align}
\widetilde{M}_{z}:&(x+1/2,y+1/2,-z)i\sigma_z; \\
P:&(-x,-y,-z)\sigma_0; \\
M_x:&(-x,y,z)i\sigma_x,
\end{align}
where $\sigma_i $~$(i=x,y,z)$ are Pauli matrices for the spin degree of freedom and  $s_0$ is the 2$\times$2 identity matrix. Here, the tilde in $\widetilde{M}_{z}$ indicates that it is a nonsymmorphic glide mirror operation---the mirror reflection is accompanied with a half lattice translation parallel to the mirror plane. We shall see that the three symmetries $\widetilde{M}_{z}$, $P$, and $T$ dictate the existence of Dirac points at $\bar{\mathrm{X}}_1$ and $\bar{\mathrm{X}}_2$, which are robust under SOC.

We first show that the three symmetries guarantee fourfold band degeneracies at $\bar{\mathrm{X}}_1$ and $\bar{\mathrm{X}}_2$. The key point is that the nonsymmorphic character of $\widetilde{M}_{z}$ leads to a special commutation relation between $\widetilde{M}_{z}$ and $P$. To see this, we compare the results when these two symmetry operators act on $(x,y,z)$ in the different orders:
\begin{small}
\begin{align}
(x,y,z)&\xrightarrow{P}(-x,-y,-z) \xrightarrow{\widetilde{M}_{z}} (-x+1/2,-y+1/2,z), \\
(x,y,z)&\xrightarrow{\widetilde{M}_{z}}(x+1/2,y+1/2,-z) \xrightarrow{P} (-x-1/2,-y-1/2,z).
\end{align}
\end{small}
This means that 
\begin{equation}\label{MP}
\widetilde{M}_{z}P=T_{110}P\widetilde{M}_{z},
\end{equation}
where $T_{110}=e^{-ik_x-ik_y}$ denotes the translation by one unit cell along both $x$ and $y$ directions. Consequently, at the special high-symmetry points $\bar{\mathrm{X}}_{1}:(\pi,0)$ and $\bar{\mathrm{X}}_{2}:(0, \pi)$, $\widetilde{M}_{z}$ and $P$ anti-commute with each other: $\{\widetilde{M}_{z},P\}=0$. Meanwhile, the nonsymmorphic character also makes the eigenvalues of $\widetilde{M}_{z}$ momentum-dependent. Since $(\widetilde{M}_{z})^2=-T_{110}$ (the minus sign is due to a $2\pi$ rotation on spin), we have the $\widetilde{M}_{z}$ eigenvalues $g_z=\pm i e^{-ik_x/2-ik_y/2}$. Importantly, at $\bar{\mathrm{X}}_1$ and $\bar{\mathrm{X}}_2$, $g_z=\pm 1$, which are purely real.

Consider an energy eigenstate $\ket{\Phi(\bar{\mathrm{X}}_i)}$ at $\bar{\mathrm{X}}_i$ $(i=1,2)$ which can be chosen as an eigenstate of $\widetilde{M}_{z}$ with eigenvalue $g_z$. As $\bar{\mathrm{X}}_i$ is a $T$-invariant momentum point, $\ket{\Phi(\bar{\mathrm{X}}_i)}$ has a degenerate Kramers partner $T\ket{\Phi(\bar{\mathrm{X}}_i)}$, which must share the same $g_z$ eigenvalue (since $g_z=\pm 1$ is real). Moreover, because $\{\widetilde{M}_{z},P\}=0$, $P\ket{\Phi(\bar{\mathrm{X}}_i)}$ (and also $TP\ket{\Phi(\bar{\mathrm{X}}_i)}$) must be another degenerate partner of $\ket{\Phi(\bar{\mathrm{X}}_i)}$ with an \emph{opposite} $\widetilde{M}_{z}$ eigenvalue $(-g_z)$. Thus, the four states ($\ket{\Phi}$, $P\ket{\Phi}$,  $T\ket{\Phi}$, $TP\ket{\Phi}$) always form a degenerate quartet at $\bar{\mathrm{X}}_1$ and $\bar{\mathrm{X}}_2$. Deviating from $\bar{\mathrm{X}}_i$, the fourfold degeneracy will generally be lifted because the $k$ point is no longer invariant under $T$.  We note that Eq.~(6) plays the pivotal role in the formation of band degeneracy. The phase factor from $T_{110}$, reflecting the nonsymmorphic nature of the lattice, determines the existence and location of the Dirac points. In addition, the above argument is made with explicit consideration of SOC, so these Dirac points are indeed robust against SOC, and can be termed as the 2D spin-orbit Dirac points. 

To further characterize the emergent 2D spin-orbit Dirac fermions and to show that the dispersion is indeed of linear type, we construct an effective $k\cdot p$ model around each Dirac point based on the symmetry constraints. Consider DP1 at $\bar{\mathrm{X}}_1$, the symmetry operations in the little group at $\bar{\mathrm{X}}_1$ include $T$ and the three generators in Eqs.~(1-3). The matrix representations of these operators can be obtained from the standard reference~\cite{Bradley}, with $T=-i\sigma_y\otimes \tau_0 K$, $\widetilde{M}_{z}=\sigma_{z}\otimes \tau_y$, $P=\sigma_0\otimes\tau_x$, and $M_x=-i\sigma_x\otimes\tau_x$. Here, $K$ is the complex conjugation operator, $\sigma_j$ and $\tau_j$ ($j=x,y,z$) are the Pauli matrices representing spin and orbital  degrees of freedom, respectively, $\sigma_0$ and $\tau_0$ are the $2\times 2$ identity matrices. Subjected to these symmetry constraints, the effective model in the vicinity of DP1 expanded to linear order in the wave vector $k'$ takes the form of
\begin{equation}\label{Heff1}
\mathcal{H}(\bm k')=v_x k'_x(\cos\theta\ \sigma_x\otimes\tau_z+\sin\theta\ \sigma_0\otimes\tau_y)+v_yk'_y\sigma_y\otimes \tau_z,
\end{equation}
where the energy and the wave vector $\bm k'=(k'_x, k'_y)$ are measured from DP1, the model parameters $v_x$, $v_y$, and $\theta$ are real, and their values depend on the microscopic details. The dispersion around DP1 is given by $E=\pm \sqrt{v_{x}^{2} {k'}_{x}^{2}+v_{y}^{2} {k'}_{y}^{2}}$, which indeed corresponds to a linear Dirac cone. This confirms that the emergent fermions are 2D spin-orbit Dirac fermions. 

The effective model for DP2 at $\bar{\mathrm{X}}_2$ can be defined in a similar way. With $T=-i\sigma_y\otimes \tau_0 K$, $\widetilde{M}_{z}=\sigma_z\otimes \tau_y$, $P=\sigma_0\otimes\tau_x$, and $M_x=-i\sigma_x\otimes\tau_0$, the effective Hamiltonian can be written as
\begin{equation}\label{Heff2}
\mathcal{H}(\bm k')=v_x k'_x\sigma_y\otimes \tau_z+v_yk'_y(\cos\theta\ \sigma_x\otimes\tau_z+\sin\theta\ \sigma_0\otimes\tau_y),
\end{equation}
Since there is no symmetry operation connecting DP1 and DP2, the model parameters are generally different for DP1 and DP2. For example, from fitting the DFT band structure, at DP1, we find the Fermi velocities $v_x=3.95\times 10^5\ m/s$ and $v_y=2.12\times 10^5\ m/s$; while at DP2, $v_x=1.19\times 10^5\ m/s$ and $v_y=4.67\times 10^5\ m/s$. The lack of symmetry connection implies that each Dirac point may be tuned separately. For example, a single Dirac point may be tuned close to the Fermi level, by lattice deformations that preserve the crystal symmetry. This is in contrast to the case of graphene, where the two Dirac points are symmetry connected and hence must shift in energy together.

The strong spin-orbit coupling in $\alpha$-Bi is crucial for the formation of the nonsymmorphic Dirac states. To see this, we calculated the band structure without the inclusion of SOC; the result is shown in Fig.~4\textbf{c}. The bands become degenerate along $\bar{\mathrm{X}}_1$-$\bar{\mathrm{M}}$-$\bar{\mathrm{X}}_2$ and, consequently, form a nodal line at the boundary of the Brillouin zone; see Fig.~4{\bf e}. The band degeneracy is also due to the nonsymmorphic symmetry of the lattice (please see the supplementary information for details). However, this band degeneracy is not robust against spin-orbit coupling. Turning on SOC, the nodal line is gapped everywhere except $\bar{\mathrm{X}}_1$ and $\bar{\mathrm{X}}_2$. In other words, SOC transforms the system from a nodal-line system into a Dirac fermion state. In the presence of SOC, the Dirac points at $\bar{\mathrm{X}}_1$ and $\bar{\mathrm{X}}_2$ are under the protection of the glided mirror symmetry. Naturally, breaking this glided mirror symmetry will lead to energy gaps at the Dirac points. Fig.~4\textbf{d} shows the band structure of a distorted lattice. The lattice distortion is depicted in the inset of Fig.~4\textbf{d}, which destroys the the glided mirror symmetry while keeping the space inversion symmetry. In this case, every band still possesses the 2-fold Kramers degeneracy but the Dirac points at $\bar{\mathrm{X}}_1$ and $\bar{\mathrm{X}}_2$ disappear. This evolution of band surfaces and Dirac/nodal points, as schematically depicted in Fig.~4\textbf{e}, indicates that SOC and nonsymmorphic symmetry are two essential pillars supporting the formation of Dirac fermions in $\alpha$-Bi.


$\alpha$-bismuthene, a two-dimensional spin-orbit material, hosts Dirac-fermion states at the high-symmetry momentum points $\bar{\mathrm{X}}_1$ and $\bar{\mathrm{X}}_2$ as demonstrated by our micro-ARPES measurements and first-principles calculations. The band degeneracy at the Dirac points are strictly protected by the nonsymmorphic symmetry of the lattice. Unlike graphene and other known 2D Dirac materials, the nonsymmorphic symmetry guarantees that the Dirac states in $\alpha$-bismuthene are robust against spin-orbit coupling. Breaking the lattice symmetry, on the other hand, can lift the band degeneracy at Dirac points and yield gapped phases. Interestingly, a surface buckling in $\alpha$-bismuthene breaks the space inversion symmetry and turns the system into a 2D elemental ferroelectric \cite{Xiao2018}. 

The formation mechanism of Dirac bands discussed in this work is intrinsically different from the band crossing induced by nonsymmorphic crystalline symmetry reported in previous works\cite{Fang2015, Hourglass2016, Yang2018, PhysRevB.96.155206, PhysRevB.95.075135}. There two bands exchange the eigenvalues of a single nonsymmorphic operator, a glided plane or a screw axis, as they disperse from one high symmetry momentum to another high symmetry momentum. Therefore, a band crossing must happen in between the two high-symmetry points. By contrast, the Dirac points in $\alpha$-bismuthene reside at the high-symmetry momenta of the Brillouin zone, because these points are invariant under $P$ and $T$ and allow momentum-dependent commutation/anticommutation relations involving the nonsymmorphic symmetry operators. In other words, the location of the Dirac points is determined by the nonsymmorphic symmetry operations\cite{Kane2016}. This property facilitates the detection of Dirac states in experiments. For example, let us consider a different nonsymmorphic group, the \#15 layer group  ($p2_{1}/m11$). A bismuth monolayer structure belongs to this layer group, please see the supplementary information. Because of a screw axis of the lattice, Dirac states are guaranteed to exist at $\bar{\mathrm{X}}_1$ and $\bar{\mathrm{M}}$ points of the Brillouin zone.  The principle demonstrated in this work can be applied to all 2D layered materials with a lattice belonging to one of the 36 nonsymmorphic layer groups\cite{Kane2016, Po2017}. This will significantly accelerate the search of 2D Dirac materials and extend ``graphene" physics into new territory where strong spin-orbit coupling is present.

 \section{Acknowledgements}
This work was supported by the MacDiarmid Institute for Advanced Materials and Nanotechnology (S.A.B. and T.M.), the National Science Center, Poland (DEC-2015/17/B/ST3/02362, P.J.K.), the Singapore Ministry of Education Academic Research Fund Tier 2 (MOE2015-T2-2-144, S.A.Y.), the National Natural Science Foundation of China (11204133, X.X.W.), and the U.S. National Science Foundation (NSF-DMR-1305583, T.-C.C. and NSF-DMR-0054904, G.B.).

\section{Author contributions}
T.O.M., F.G and A.L. performed all LEEM, LEED and ARPES experiments, as well as the Bi growth and sample preparation, with the assistance of S.B., T.M. and P.J.K.; P.J.K. and I.Z. did LEED analyses; Q.L. and G.B. performed STM measurements; X.W., T.-C.C., and G.B. performed first-principles calculations; C.-K.C., S.A.Y., Y.L., and G.B. did theoretical analyses;  P.J.K., S.A.B. and G.B. designed the project and wrote the manuscript. All authors discussed the manuscript.

\section{Correspondence}
Correspondence and requests for materials should be addressed to G.B.~(biang@missouri.edu), P.J.K. (pawel.kowalczyk@uni.lodz.pl) and S.A.B. (simon.brown@canterbury.ac.nz).

\newpage
\section{}

\newpage
\section{}

\newpage
\section{}
\begin{figure}
\includegraphics[width=1\linewidth]{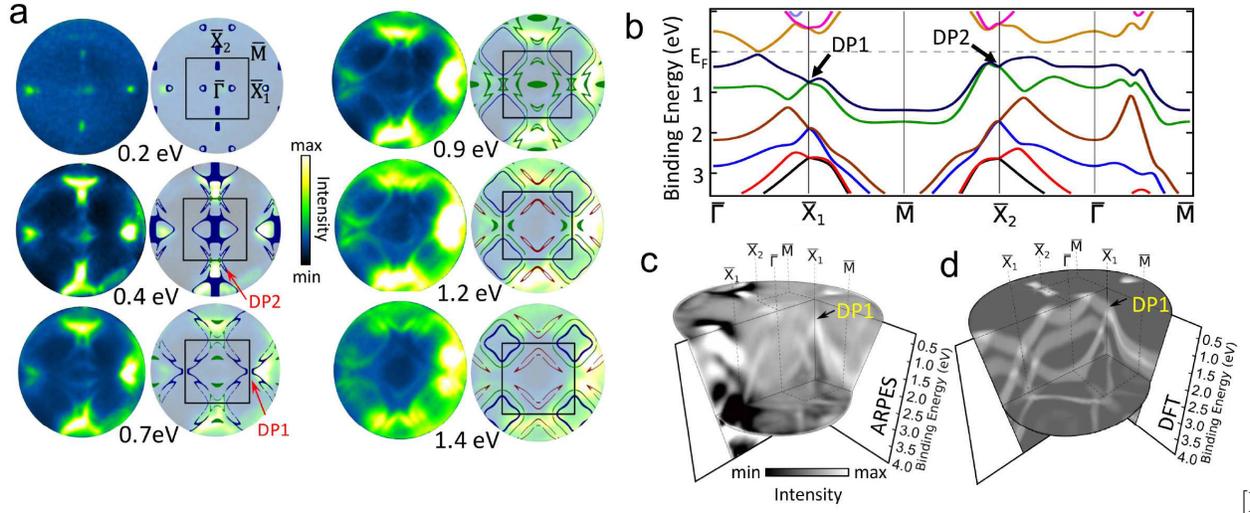}[H]
\caption{
\label{fig:mdc}
\textbf{$\mu$-ARPES iso-energy contours and first-principles simulation.} \textbf{a,} The $\mu$-ARPES iso-energy contours taken at binding energy of 0.2, 0.4, 0.7, 0.9, 1.2, and 1.4 eV. \textbf{b,} DFT calculated bands showing the location of Dirac points. \textbf{c,} 3D band representation with two cross section planes $\bar{\mathrm{X}}_1$-M-$\bar{\mathrm{X}}_1$ and $\bar{\mathrm{X}}_1$-$\bar{\mathrm{\Gamma}}$-$\bar{\mathrm{X}}_1$. \textbf{d,} Calculated 3D band contour with smearing of 0.5~eV.}%
\end{figure}

\newpage
\section{}

\begin{figure}
\includegraphics[width=1\linewidth]{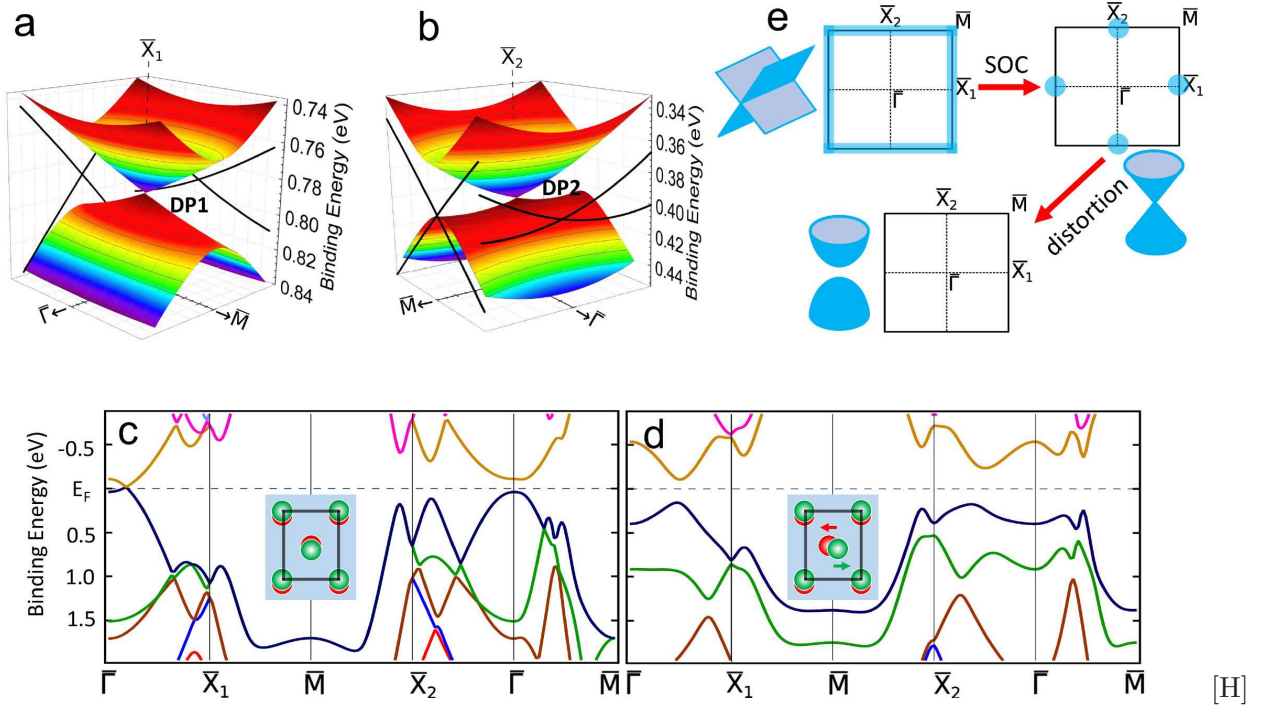}[H]
\caption{
\label{fig:dirac}
\textbf{SOC and distortion effects on $\mathbf {\alpha }$-Bi band structure.} \textbf{a, b} DFT band surface calculated in the vicinity of $\bar{\mathrm{X}}_1$ and $\bar{\mathrm{X}}_2$, respectively. \textbf{c,} Band structure of $\mathbf{\alpha}$-Bi in the absence of spin-orbit coupling. \textbf{d,} Band structure of distorted $\mathbf{\alpha}$-Bi. The distortion is shown in the inset. The two atoms are moved along arrow directions by 2\% of the unit-cell width. \textbf{e,} The evolution of band configuration from without SOC to with SOC to with both SOC and symmetry-breaking distortion. The Dirac/nodal points are highlighted in blue in the Brillouin zone.}%
\end{figure}

\newpage

\bibliographystyle{naturemag}
\bibliography{Bi_post}

\end{document}